\documentclass[12pt,preprint]{aastex}

\shorttitle{XRF020903}
\shortauthors{Soderberg et al.}

\begin{document}


\title{A Redshift Determination for XRF 020903: First Spectroscopic Observations of an X-Ray Flash}

\author{A. M. Soderberg, S. R. Kulkarni, E. Berger, D. B. Fox, P. A. Price, S. A. Yost and M. P. Hunt}
\affil{Division of Physics, Mathematics and Astronomy, MS 105-24, California Institute of Technology, Pasadena, CA 91125}

\author{D. A. Frail and R. C. Walker}
\affil{National Radio Astronomy Observatory, P.O. Box 0, Socorro, NM 87801}

\author{M. Hamuy and S. A. Shectman}
\affil{Carnegie Observatories, 813 Santa Barbara Street, Pasadena, CA 91101}

\and

\author{J. P. Halpern and N. Mirabal}
\affil{Astronomy Department, Mailcode 5246,  Columbia University, New York, NY 10027}

\begin{abstract}
We report the discovery of optical and radio afterglow emission from
the extremely soft X-ray flash, XRF 020903.  Our spectroscopic
observations provide the first redshift for an X-ray flash, thereby
setting the distance scale for these events.  At $z=0.251$, XRF 020903
is one of the nearest cosmic explosions ever detected, second only to
the recent GRB 030329 and the unusual GRB 980425/SN 1998bw.  Moreover,
XRF 020903 is the first X-ray flash for which we detect an optical
afterglow.
The luminosity of the radio afterglow of XRF 020903 is 1000 times
greater than that of Ibc supernovae but similar to those of GRB afterglows.
From broadband afterglow modeling we show that the explosion energy
of XRF 020903 is not dissimilar from values inferred for typical
gamma-ray bursts, suggesting that these cosmological explosions may
derive from a similar mechanism.  

\end{abstract}


\section{Introduction}
Prior to the detection of afterglows, gamma-ray bursts (GRBs) were
enshrouded in mystery for nearly thirty years.  Great progress in our
understanding of these energetic events came with the first redshift
measurement which placed GRBs at cosmological distances
\citep{mdk+97}.  In a similar fashion, the mystery of X-ray flashes
(XRFs) has been fueled by the absence of confirmed redshifts.  These
events, identified in the 1990s by {\em BeppoSAX}, are characterized
by a peak energy in $\nu F_{\nu}$ of $E_{\rm peak}\sim 25$ keV
(compared to $E_{\rm peak}\sim 250$ keV for GRBs).  With a
distribution of durations similar to those observed for GRBs, it has
been assumed that XRFs are associated with GRBs and therefore share
their extragalactic distance scale \citep{hzk+01}.

The subsequent discovery of XRF X-ray and radio afterglows with
properties similar to those observed in GRB afterglows has further
strengthened this association \citep{hyf+01, acf+02, tfk01}.  Still,
the question of whether the difference between GRBs and XRFs is
intrinsic or extrinsic remains unanswered.  Assuming XRFs are simply
GRBs observed away from the jet collimation axis, they would have less
$\gamma$-ray emission and the difference is extrinsic, based solely on
the line-of-sight to the observer.  On the other hand, the difference
could be intrinsic, namely XRFs may represent a class of explosions
which are similar in energetics to GRBs yet characterized by less
relativistic ejecta possibly due to a heavier baryonic load.  It is
clear that by setting the distance scale for XRFs (and hence their energy 
scale) we can begin to distinguish between extrinsic and
intrinsic effects.

In this paper we present the first spectroscopic redshift for
an X-ray flash, XRF 020903, placing this event among the nearest
high energy explosions and offering confirmation that
X-ray flashes are cosmological and produce a total energy output 
similar to that observed in GRBs.

\section{Observations}
On 2002 September 3.421 UT the Wide-Field X-ray Monitor (WXM) and Soft X-ray
Camera (SXC) aboard the High Energy Transient Explorer-2 (HETE-2)
detected an X-ray flash within the 0.5-10 keV energy band.  With an
exceptionally low peak energy of $E_{\rm peak}\sim 5$ keV and a
fluence of $7.2\times 10^{-8}~\rm erg~cm^{-2}$, XRF 020903 is the
softest event ever detected by HETE-2 with a ratio of
X-ray fluence ($S_{X}$) to $\gamma$-ray fluence ($S_{\gamma}$) of
log($S_{X}/S_{\gamma})=4.3$ \citep{slg+03}.  Ground analysis provided
a localization for XRF 020903 centered at $\alpha\rm (J2000)=22^{\rm
h}49^{\rm m}01^{\rm s}$, $\delta\rm (J2000)=-20^{\rm o}55'47''$
with a $4'\times 31'$ uncertainty region at $\Delta t\approx 0.3$
days \citep{rak+02}.

\subsection{Ground-based Photometry}
We began observing the field of XRF 020903 on 2002 September 4.32 UT
($\Delta t\approx 0.9$ days) with the Palomar Observatory 200-inch
telescope (P200) equipped with the Large Field Camera (LFC).  With a
total exposure time of 20 minutes under photometric conditions
(stellar $\rm FWHM\sim 1.2''$) the observations reached a limiting
magnitude of $R\sim 23$ mag.  Visual comparison with Digitized Sky
Survey archival images did not reveal an afterglow candidate.  

\begin{figure}
\plotone{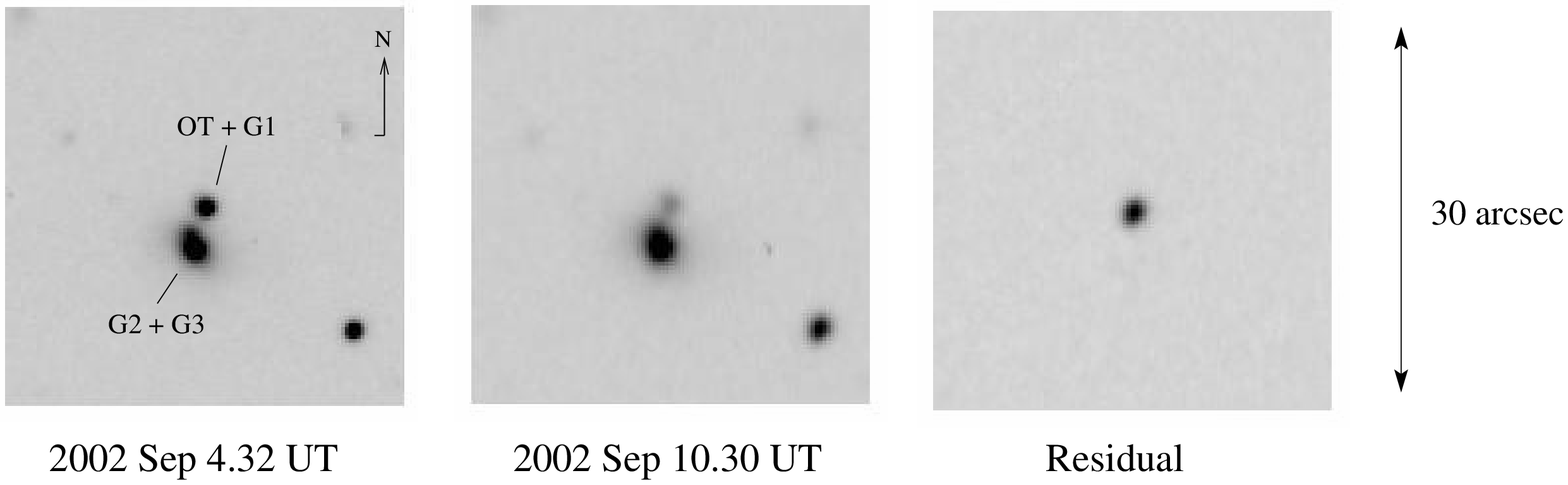}
\caption{The field of XRF 020903 was observed with the Palomar 200
inch telescope equipped with the Large Field Camera on 2002 September
4.32 and September 10.30 UT.  Image subtraction techniques revealed
one variable object within the HETE-2 error-box lying $\sim4$ arcsec
NW of a bright elliptical galaxy (G2).  The residual image clearly
indicates a transient stellar source which decreased in brightness by
1.4 magnitudes between the two epochs.}
\label{fig:XRF020903-field}
\end{figure}

A second
epoch was obtained on September 10.30 UT ($\Delta t\approx$7 days)
using the same observational set-up and in similar observing
conditions. 
Image subtraction between the first and second epochs
revealed one variable object within the HETE-2 error region 
(Figure~\ref{fig:XRF020903-field})
located at $\alpha\rm (J2000)=22^{\rm h}48^{\rm m}42.34^{\rm s}$,
$\delta\rm (J2000)-20^{\rm o}46'09.3''$ and lying $\sim4$ arcsec NW of
a bright elliptical galaxy (hereafter G2; \citealt{spf+02}).

With an approximate magnitude of $R\approx 19$ at $\Delta t\approx 1$ day, the new
object decreased in brightness by 1.4 magnitudes between the two
epochs, implying a temporal flux decay index of $\alpha\approx -1$.
As the source proved consistent with a typical GRB afterglow
evolution, the optical transient was adopted as a suitable candidate
for the optical afterglow of XRF 020903.

\begin{figure}
\plotone{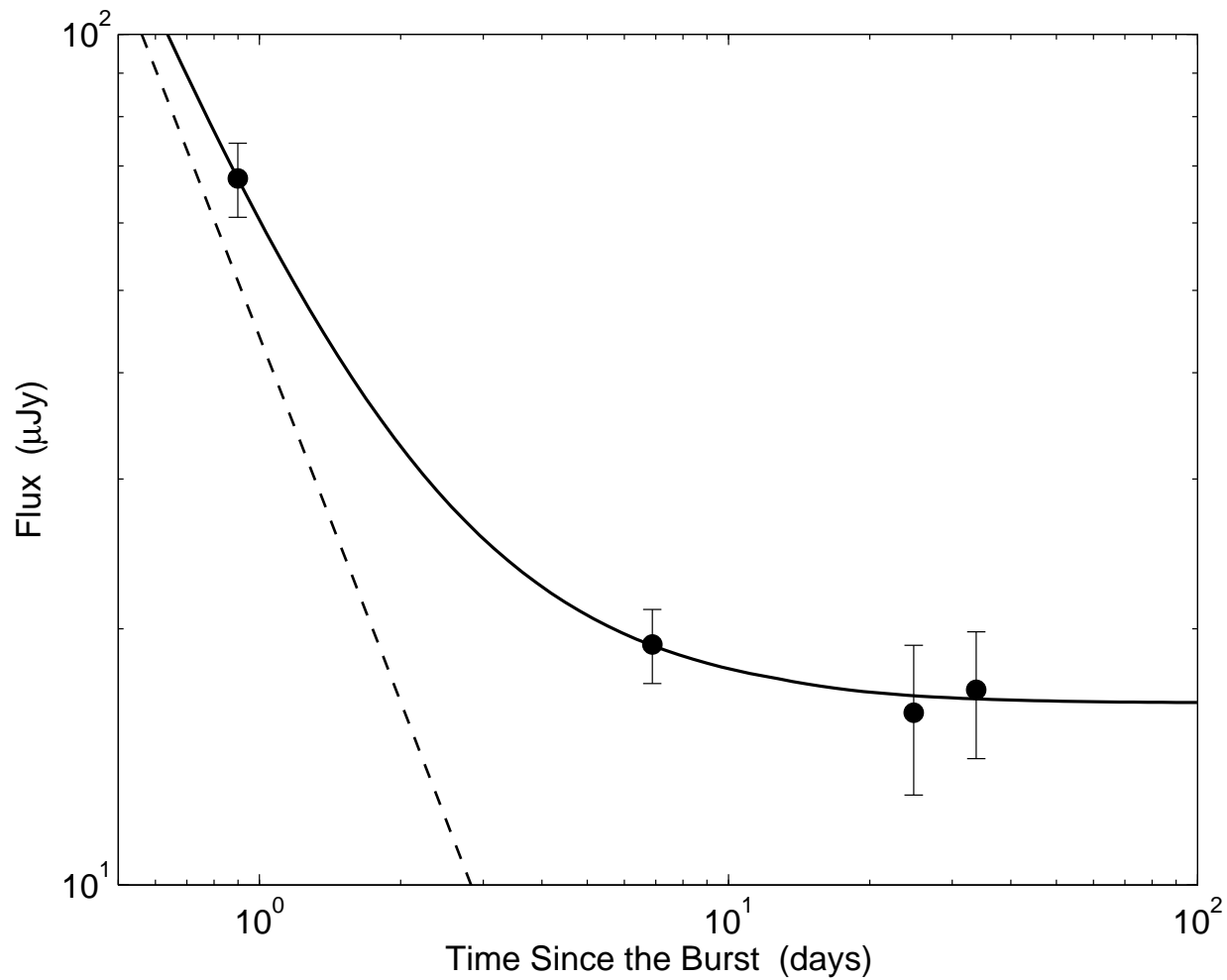}
\caption{Optical (R band) lightcurve for the transient source. Data
were obtained from the Palomar 200 inch and MDM 1.3 meter
telescopes. 
The solid line is our best fit to the broadband data, including
the afterglow component with a temporal decay of $\alpha \approx -1.3$
(dashed line) and the host galaxy (G1) contribution of $R\sim 20.8$ mag.
The early decline rules out the possibility that the viewing angle
is significantly away from the jet axis (see \citealt{gpk+02}). }
\label{fig:XRF020903-Optical}
\end{figure}

We observed the afterglow position on three additional epochs with the
P200 and the MDM Observatory 1.3 meter telescope (Table~\ref{tab:Optical-Photometry} and Figure~\ref{fig:XRF020903-Optical}).
These, along with Digitized Sky Survey archival images, reveal
the presence of an extended source with $R\sim 21$ mag
coincident with the position of the transient and thereby suggestive
of a host galaxy (hereafter G1).

Due to the underlying extended source and the proximity of
the transient to G2, accurate magnitude estimates relied on careful
PSF photometry of the field.  Absolute calibration of field stars was
supplied by \citet{h02}.  We used 12 unsaturated field stars in common
between the Palomar and Henden images to fully calibrate observations
of the variable source.  
We note that although the transient decreased in
brightness quickly between the first two epochs, later epochs
($\Delta t\approx 30-40$ days) appear to indicate a plateau in the light curve
which was confirmed by other observers \citep{fap+02} and may
originate from unresolved flux contamination from G1.

\subsection{Ground-based Spectroscopy}
Initial spectroscopy of the transient was performed on 2002 September 28.1 UT
with the Magellan Baade Telescope using the Low Dispersion Survey
Spectrograph (LDSS2).  A position angle of 168 degrees was used such
that spectral information on G1 and G2 was obtained simultaneously.
Despite the positional coincidence of G2 and G1, it was found that the
systems are not physically associated, separated by 3900 km/s in
velocity space \citep{spf+02}.  In this epoch, the transient source was still
significantly bright.

Further spectroscopic observations were made of the putative host (G1)
with the Echelle Spectrograph $\&$ Imager (ESI) mounted on the Keck II
telescope on 2003 July 4 UT ($\Delta t\approx 300$ days).  
These observations do not include
any flux from the transient source.  During a total
exposure time of 1 hour, we obtained a spectrum of G1 and G2 with
a slit width of 0.75 arcsec.  

We find that G2 is a large elliptical galaxy with at least one
interacting galaxy companion, G3, located $< 1$'' to the NE
(see Figure~\ref{fig:XRF020903-HST}).  Observations of G2 exhibit a
relatively smooth continuum with features typical of an elliptical
galaxy. The Ca II H and K absorption lines give a redshift of
$z=0.235$, while G3 is offset by only 240 km/s at a redshift of
$z=0.236$.

In contrast, G1 is shown to be an active star-forming galaxy at z=0.251
with a rich set of narrow bright emission lines (Figure 2).  The
[OIII]/H$\beta$ and [NII]/H$\alpha$ intensity ratios indicate that the
galaxy is a low-metallicity and high-excitation starburst galaxy.  In
addition, the flux ratio of the [NeIII] and [OII] lines is
$F^{3869}/F^{3727}\approx 0.43$, similar to the observed value for the
host galaxy of GRB 970508 \citep{bdk+98} and approximately 10 times
higher than typical values for H II regions.  The bright [NeIII]
emission lines observed in GRB hosts are thought to be indicative of a
substantial population of massive stars.  On the other hand,
\citet{cf02} note that a spectrum of G1 taken with the Low Resolution
Imaging Spectrometer (LRIS) on the Keck I telescope reveals a deficit
of emission at rest wavelengths $< 4000 \rm \AA$ which is consistent
with a population of older stars.

\subsection{Hubble Space Telescope}
The afterglow candidate was observed with the Hubble Space Telescope
(HST) using the Advanced Camera for Surveys (ACS) under Program No.9405
(P.I.: A. Fruchter).  Three epochs of imaging were obtained from
$\Delta t\sim 94$ to $\sim 300$ days with exposure times of 1840 sec in
the F606W filter. Following ``On-The-Fly'' pre-processing the data were
drizzled using standard IRAF tools (STSDAS; \citealt{fh02}).  In
drizzling our final images, we retained the native WFC pixel scale of
0.05'' and used a {\tt pixfrac} of 1.0.  The HST images reveal a
complex galaxy morphology for G1, suggesting a system of at least four
interacting galaxies (see Figure~\ref{fig:XRF020903-HST}).

To locate the optical transient with respect to the host galaxy
complex, we performed a source to source comparison of our first-epoch
Palomar (2002 September 4) and HST (2002 December 3) images.  We found
42 unsaturated, unconfused sources in common between these two images,
and were able to match the two coordinate lists with an rms mapping
uncertainty of 0.06~arcsec.  We derived the position of the transient
from the difference image of the 2002 September 4 and September 10
data.  Since the optical transient is well-detected in the difference
image, the uncertainty in its centroid position is negligible.  The
uncertainty in the coordinate mapping thus dominates the uncertainty
in the position of the source relative to the host galaxy complex.
The optical transient appears to overlap with the SW
component of the G1 complex (Figure~\ref{fig:XRF020903-HST}).

\subsection{Radio Observations of XRF 020903}
\subsubsection{Very Large Array Data} 
We began observations of the field of XRF 020903 with the Very Large
Array (VLA\footnotemark\footnotetext{The VLA is operated by the National
Radio Astronomy Observatory, a facility of the National Science
Foundation operated under cooperative agreement by Associated
Universities, Inc.}) on 2002 September 27.22 UT.  A radio source was detected
in coincidence with the optical transient at a location of $\alpha\rm
(J2000)=22^{\rm h}48^{\rm m}42.34^{\rm s}, \delta\rm (J2000)=-20^{\rm
o}46'08.9''$ with an uncertainty of 0.1 arcsec in each coordinate.
The initial observation showed the radio source to have a flux density
of $F_{\nu}=1.06\pm0.02$ mJy at 8.5 GHz.  The National Radio Astronomy
Observatory VLA Sky Survey (NVSS; \citealt{ccg+98}) did not show any
evidence for a pre-existing source at this location down to a limit of
1 mJy.  Further observations at 8.5 GHz on September 29.11 ($\Delta
t\approx 26$ days) showed that the source faded to
$F_{\nu}=0.75\pm0.04$ mJy.  We continued monitoring the transient
source with the VLA over the next $\approx 370$ days at frequencies of
1.5, 4.9, 8.5 and 22.5 GHz.  The lightcurve is displayed in
Figure~\ref{fig:XRF020903-Radio}.

\subsubsection{Very Long Baseline Array Data}
The relatively low redshift and strong radio emission of the transient
source made it an ideal candidate for Very Long Baseline Array (VLBA)
observations.  On September 30.23 UT ($\Delta t\approx 27$ days) we
observed the radio transient for a total duration of 6 hours at 8.5
GHz.  Using sources J2253+1608 and J2148+0657 we were able to flux
calibrate the field and we phase referenced against source J2256-2011
at a distance of $<2$ degrees.  Data were reduced and processed using
standard VLBA packages within the Astronomical Image Processing System
(AIPS).  We detected the radio transient with a flux of $0.89\pm0.17$
mJy at a position coincident with the VLA and optical observations at
a location of $\alpha\rm (J2000)=22^{\rm h}48^{\rm m}42.33912^{\rm
s}\pm0.00003$, $\delta\rm (J2000)=-20^{\rm o}46'08.945''\pm 0.0005$.
This is our most accurate position measurement for the transient object.
The source is unresolved within our VLBA beam size of $1.93\times
1.73$ mas.  Furthermore, the consistency between the flux measured with
VLBA and low resolution VLA observations rules out the presence of
diffuse components (e.g. jets).  

\section{The afterglow of XRF 020903}

The large error box and the proximity of the optical transient to G2
and G3 delayed rapid identification of the transient
\citep{psa02,prp02,uik+02,fsm+02}.  As a result there was little
optical followup of the transient at early times (see
Table~\ref{tab:Optical-Photometry}).  At later times, the transient is
rapidly dominated by the emission from G1.  An additional complication
was introduced by \citet{g02} who, based on archival photographic
plates from 1954 and 1977, proposed that G1 hosted a variable active
galactic nucleus (AGN).

In contrast to the above discussion, the extensive radio light curve
(Figure~\ref{fig:XRF020903-Radio}) provides strong evidence that the
transient is the afterglow of XRF 020903.  The radio light curve is
similar to the afterglow of GRBs \citep{fkb+03}.  At 4.9 GHz, the
source was observed to rise to a peak flux at $\Delta t\approx29$ days
and subsequently decay with a characteristic index of $\alpha\approx
-1.1$.  A steeper decline of $F_{\nu}\propto t^{-1.5}$ was observed at
8.5 GHz where the peak flux precedes the first observational epoch.
We note that the dynamic range of the radio afterglow emission clearly
distinguishes the source from a radio bright AGN.

The afterglow interpretation is consistent with the optical
spectroscopic observations of G1, namely the absence of broad lines or
features typical of AGN, as well as the absence of any jet
structure in the VLBA images of the radio transient. We proceed
accepting the notion that the optical/radio transient is the afterglow
of XRF 020903.

\section{Energetics}
At a redshift of $z=0.251$, the X-ray fluence implies an isotropic
equivalent energy $E_{\rm iso}\approx 1.1\times 10^{49}$ erg.  This
value is 3 to 6 orders of magnitude lower than the isotropic energies
of gamma-ray bursts \citep{fks+01}.  Since there is no evidence for a
jet break in the afterglow observations, the data are consistent with
a wide jet or a spherical explosion.  This, along with the early
decline of the optical flux ($F_{\nu}\propto t^{-1.3}$ at $\Delta
t\approx 1$ d; Figure~\ref{fig:XRF020903-Optical}) rules out the
possibility that the inferred low energy is due to a viewing angle
significantly away from the jet axis.  In such a scenario, the optical
flux would rise until the edge of the jet enters the observer's line
of sight, peaking at $\Delta t \ge t_{jet}$, where $t_{jet}$ is the
observed time of the jet break (e.g. \citealt{gpk+02}).

Figure~\ref{fig:XRF020903-GRB-radio-hist}  
shows that although the energy in the prompt emission of XRF
020903 is significantly lower than that observed for typical GRBs, the
peak luminosity of the radio afterglow is comparable to that found in
GRB afterglows.  This indicates that XRF 020903 has a similar kinetic
energy to GRB afterglows.  To study this in more detail we used
standard broadband afterglow models with spherical and collimated
ejecta expanding into circumburst media with uniform density and wind
density profile, $\rho\propto r^{-2}$ (e.g. \citealt{bsf+00}).  We find that
independent of the assumed model, the total kinetic energy is $E\sim
4\times 10^{50}$ erg with reasonable values
for the circumburst density and energy fractions (electron and
magnetic field) of $n\approx 100~\rm cm^{-3}$, $\epsilon_e\approx 0.6$ and
$\epsilon_B\approx 0.01$, respectively.  Thus we confirm the
afterglow energy is similar to values found for GRB afterglows
\citep{pk02,bkf03}.

Recent evidence suggests a standard total energy yield ($10^{51}$ erg)
for all GRBs \citep{bkp+03}, where the total energy yield ($E_{\rm
tot}$) is defined as the sum of the energy in the prompt emission
($E_{\gamma}$) plus the mildly relativistic energy as inferred from
the afterglow ($E_{\rm rad}$).  Clearly, for XRF 020903 the explosion
energy is dominated by the mildly relativistic afterglow, while a
minor fraction of energy couples to the high Lorentz factors
characterizing the prompt emission.

Figure~\ref{fig:XRF020903-GRB-radio-hist} also compares the radio luminosity of
XRF 020903 with other spherical explosions -- type Ibc supernovae.
With a radio luminosity $\sim 1000$ times greater than typical Ibc
SNe, it is clear that XRF 020903 is a significantly more energetic
explosion.  This suggests that despite their similar explosion geometries,
XRF 020903 and type Ibc supernovae are intrinsically different -- 
possibly due to the presense of a central engine \citep{bkf+03}.

\section{Conclusions}
We present radio and optical observations of the afterglow of
XRF 020903 -- the first X-ray flash to have a
detected optical afterglow and a spectroscopic distance determination.
At a redshift of $z=0.251$, this burst has set the distance scale
for XRFs and confirmed the assumption that they are cosmological in origin.
The host galaxy of XRF 020903 appears to be a typical star-forming
galaxy similar to those of GRB host galaxies.

The isotropic energy release of the prompt (X-ray) emission of XRF
020903 is at least two orders of magnitude smaller than those of GRBs.
However, from our broadband modeling of the afterglow we determine that
the total kinetic energy is about $10^{50}\,$erg, not dissimilar to
those inferred for GRBs \citep{pk02,bkf03}.  In
comparison with a larger sample of cosmic explosions, XRF 020903 is a
clear example where less energy is coupled to high Lorentz factors.
This source highlights the diversity in high energy transients and
underscores the importance of studying spherical explosions.

\acknowledgments

We thank J. Ulvestad for his help with scheduling the VLBA
observation.  MH acknowldges support for this work provided by NASA
through Hubble Fellowship grant HST-HF-01139.01-A awarded by the Space
Telescope Science Institute, which is operated by the Association of
Universities for Research in Astronomy, Inc., for NASA, under contract
NAS 5-26555.  JPH and NM acknowledge support from NSF grant
AST-0206051. AMS is supported by an NSF graduate research
fellowship. GRB research at Caltech is supported by NASA and NSF.


\begin{thebibliography}{}

\bibitem[{Amati} {\it et al.}\ (2002)]{acf+02}
{Amati}, L., {Capalbi}, M., {Frontera}, F., {Gandolfi}, G., {Piro}, L., {in't
  Zand}, J.~J.~M., {Granata}, S., and {Reali}, F. 2002, GRB Circular Network,
  1386, 1.

\bibitem[{Berger}, {Kulkarni} \& {Frail}(2003)]{bkf03}
{Berger}, E., {Kulkarni}, S.~R., and {Frail}, D.~A. 2003, \apj, 590, 379.

\bibitem[{Berger} {\it et al.}\ (2003a)]{bkf+03}
{Berger}, E., {Kulkarni}, S.~R., {Frail}, D.~A., and {Soderberg}, A.~M. 2003a,
  \apj, 599, 408.

\bibitem[{Berger} {\it et al.}\ (2003b)]{bkp+03}
{Berger}, E. {\it et al.}\  2003b, \nat, 426, 154.

\bibitem[{Berger} {\it et al.}\ (2000)]{bsf+00}
{Berger}, E. {\it et al.}\  2000, \apj, 545, 56.

\bibitem[{Bloom} {\it et al.}\ (1998)]{bdk+98}
{Bloom}, J.~S., {Djorgovski}, S.~G., {Kulkarni}, S.~R., and {Frail}, D.~A.
  1998, \apjl, 507, L25.

\bibitem[{Chornock} \& {Filippenko}(2002)]{cf02}
{Chornock}, R. and {Filippenko}, A.~V. 2002, GRB Circular Network, 1609, 1.

\bibitem[{Condon} {\it et al.}\ (1998)]{ccg+98}
{Condon}, J.~J., {Cotton}, W.~D., {Greisen}, E.~W., {Yin}, Q.~F., {Perley},
  R.~A., {Taylor}, G.~B., and {Broderick}, J.~J. 1998, \aj, 115, 1693.

\bibitem[{Frail} {\it et al.}\ (2003)]{fkb+03}
{Frail}, D.~A., {Kulkarni}, S.~R., {Berger}, E., and {Wieringa}, M.~H. 2003,
  \aj, 125, 2299.

\bibitem[{Frail} {\it et al.}\ (2001)]{fks+01}
{Frail}, D.~A. {\it et al.}\  2001, \apjl, 562, L55.

\bibitem[{Fruchter} {\it et al.}\ (2002)]{fsm+02}
{Fruchter}, A., {Strolger}, L., {Mobasher}, B., {Rhoads}, J., {Levan}, A.,
  {Burud}, I., and {Becker}, A. 2002, GRB Circular Network, 1557, 1.

\bibitem[{Fruchter} \& {Hook}(2002)]{fh02}
{Fruchter}, A.~S. and {Hook}, R.~N. 2002, \pasp, 114, 144.

\bibitem[{Gal-Yam}(2002)]{g02}
{Gal-Yam}, A. 2002, GRB Circular Network, 1556, 1.

\bibitem[{Gorosabel} {\it et al.}\ (2002)]{fap+02}
{Gorosabel}, J., {Hjorth}, J., {Pedersen}, H., {Jensen}, B.~L., {Fynbo},
  J.~P.~U., {Andersen}, M., {Castro Ceron}, J.~M., and {Castro-Tirado}, A.~J.
  2002, GRB Circular Network, 1631, 1.

\bibitem[{Granot} {\it et al.}\ (2002)]{gpk+02}
{Granot}, J. and {Panaitescu}, A. and {Kumar}, P. and {Woosley}, S.~E.
  2002, \apj, 570, L61.

\bibitem[{Harrison} {\it et al.}\ (2001)]{hyf+01}
{Harrison}, F.~A., {Yost}, S., {Fox}, D., {Heise}, J., {Kulkarni}, S.~R.,
  {Price}, P.~A., and {Berger}, E. 2001, GRB Circular Network, 1143, 1.

\bibitem[{Heise} {\it et al.}\ (2001)]{hzk+01}
{Heise}, J., {in't Zand}, J., {Kippen}, R.~M., and {Woods}, P.~M. 2001, in {
  Gamma-ray Bursts in the Afterglow Era}, 16.

\bibitem[{Henden}(2002)]{h02}
{Henden}, A. 2002, GRB Circular Network, 1571, 1.

\bibitem[{Metzger} {\it et al.}\ (1997)]{mdk+97}
{Metzger}, M.~R., {Djorgovski}, S.~G., {Kulkarni}, S.~R., {Steidel}, C.~C.,
  {Adelberger}, K.~L., {Frail}, D.~A., {Costa}, E., and {Frontera}, F. 1997,
  \nat, 387, 878.

\bibitem[{Panaitescu} \& {Kumar}(2002)]{pk02}
{Panaitescu}, A. and {Kumar}, P. 2002, ApJ, 571, 779.

\bibitem[{Pavlenko}, {Rumyantsev} \& {Pozanenko}(2002)]{prp02}
{Pavlenko}, E., {Rumyantsev}, V., and {Pozanenko}, A. 2002, GRB Circular
  Network, 1535, 1.

\bibitem[{Price}, {Schmidt} \& {Axelrod}(2002)]{psa02}
{Price}, P.~A., {Schmidt}, B.~P., and {Axelrod}, T.~S. 2002, GRB Circular
  Network, 1533, 1.

\bibitem[{Ricker} {\it et al.}\ (2002)]{rak+02}
{Ricker}, G. {\it et al.}\  2002, GRB Circular Network, 1530, 1.

\bibitem[{Sakamoto} {\it et al.}\ (2003)]{slg+03}
{Sakamoto}, T. {\it et al.}\  2003, Submitted to ApJ, astro-ph/0309455.

\bibitem[{Soderberg} {\it et al.}\ (2002)]{spf+02}
{Soderberg}, A.~M. {\it et al.}\  2002, GRB Circular Network, 1554, 1.

\bibitem[{Taylor}, {Frail} \& {Kulkarni}(2001)]{tfk01}
{Taylor}, G.~B., {Frail}, D.~A., and {Kulkarni}, S.~R. 2001, GRB Circular
  Network, 1136, 1.

\bibitem[{Uemura} {\it et al.}\ (2002)]{uik+02}
{Uemura}, M., {Ishioka}, R., {Kato}, T., and {Yamaoka}, H. 2002, GRB Circular
  Network, 1537, 1.

\bibitem[{Yun} \& {Carilli}(2002)]{yc02}
{Yun}, M.~S. and {Carilli}, C.~L. 2002, \apj, 568, 88.

\end{thebibliography}

\clearpage 

\clearpage

\begin{figure}
\plotone{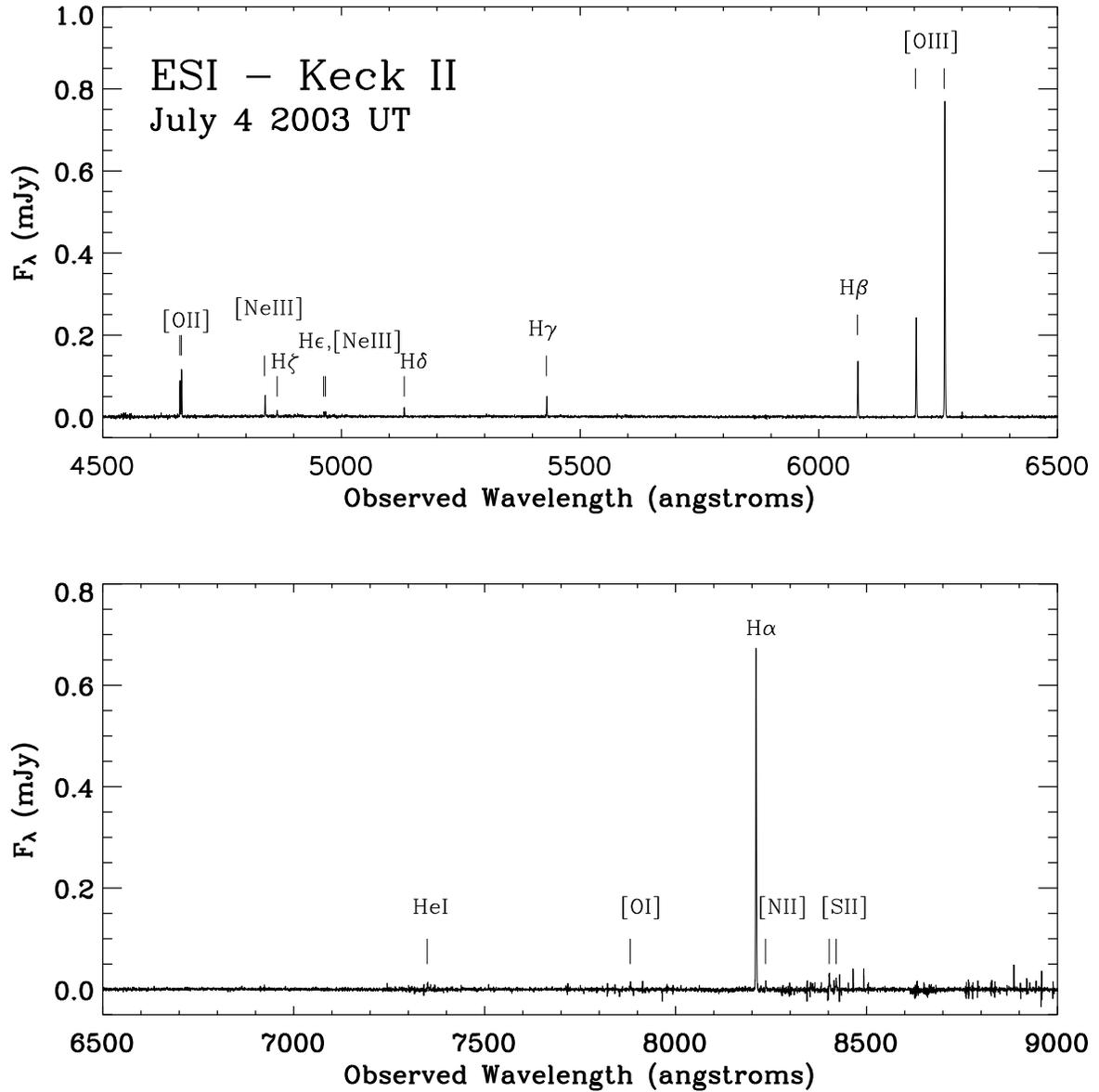}
\caption{On 2003 July 4 UT we obtained spectroscopic observations of
the putative host galaxy (G1) underlying the optical transient source
associated with XRF 020903.  Data were taken with the
Echelle Spectroscopic Imager (ESI) mounted on the Keck II
telescope. The source is shown to be an active star-forming galaxy at
z=0.251 with a rich set of narrow bright emission lines.}
\label{fig:XRF020903-Spectrum}
\end{figure}

\clearpage

\begin{figure}
\plotone{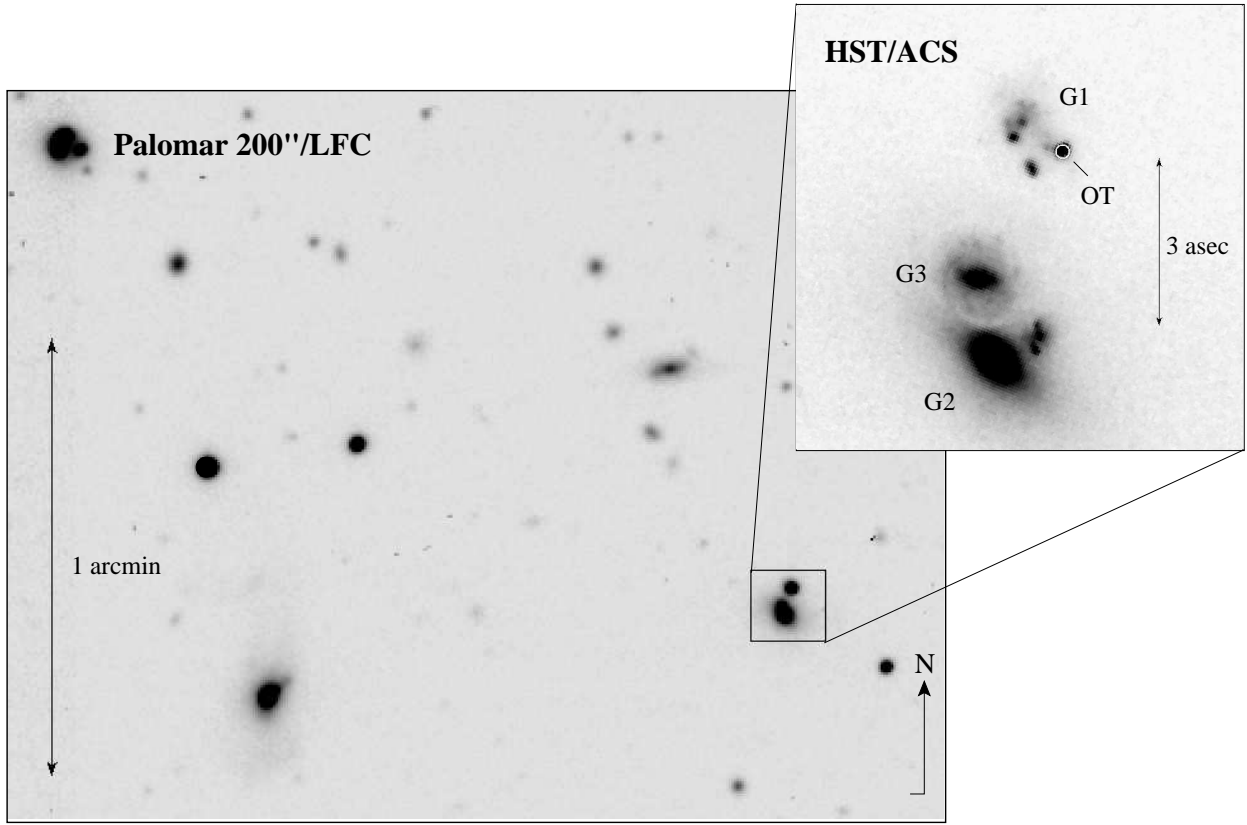}
\caption{The transient discovered within the error-box of XRF 020903
was observed with the Hubble Space Telescope (HST) using the Advanced
Camera for Surveys (ACS) on 2002 December 3 UT. The HST image reveals
a complicated galaxy morphology for G1, suggesting a system of at
least four interacting galaxies. The location of the optical transient
is noted on the $7\times7''$ HST cutout with a circle which represents
a 2$\sigma$ positional uncertainty of 0.12''.  Nearby galaxies G2 and
G3 are labeled accordingly on the HST image. The optical transient
appears to overlap with the SW component of the G1 complex.}
\label{fig:XRF020903-HST}
\end{figure}

\clearpage

\begin{figure}
\plotone{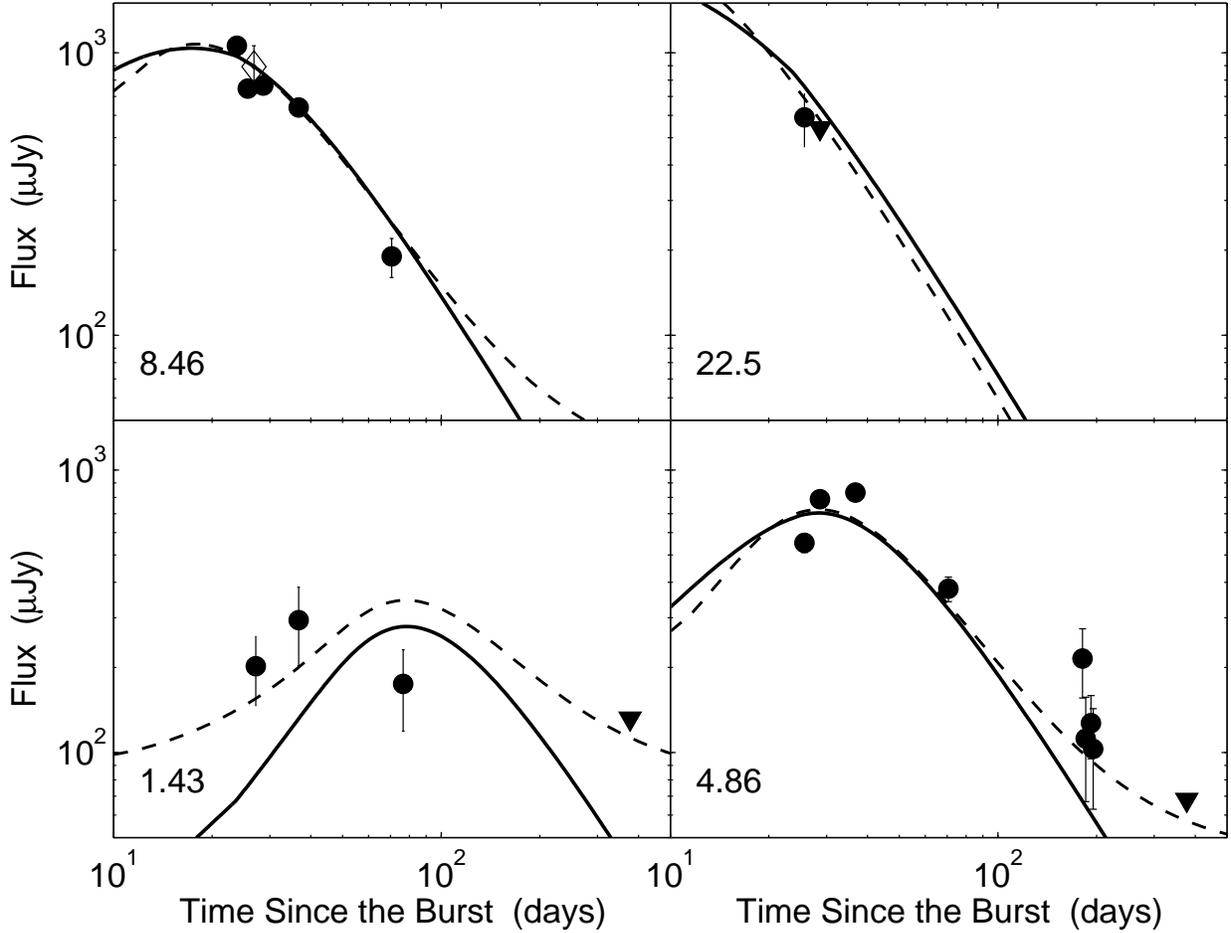}
\caption{We observed the field of XRF 020903 with the Very Large
Array over the period $\Delta t\approx 25-370$ days.
A bright radio counterpart was detected at a position coincident with the
optical transient.  We continued monitoring the source with the
VLA over the next $\approx 300$ days at frequencies of 1.5, 4.9, 8.5
and 22.5 GHz.  Assuming the transient source is the radio afterglow component
associated with XRF 020903, we overplot the best fit broadband model
(solid line) which predicts a total kinetic energy of $4\times
10^{50}$ erg in the afterglow.  Allowing for a host galaxy emission
component, we find a better fit assuming a host galaxy flux of $F_{\rm
1.4 GHz}\approx 80 \mu$Jy (dashed line).  The implied star-formation rate from
the host is 5 $M_{\odot} \rm yr^{-1}$ following the conversion from 
 \citet{yc02}.  The 8.5 GHz data point
marked by the open diamond symbol denotes the VLBA observation of the
unresolved transient source (beam size=$1.93\times 1.73$
mas).}
\label{fig:XRF020903-Radio}
\end{figure}

\clearpage

\begin{figure}
\plotone{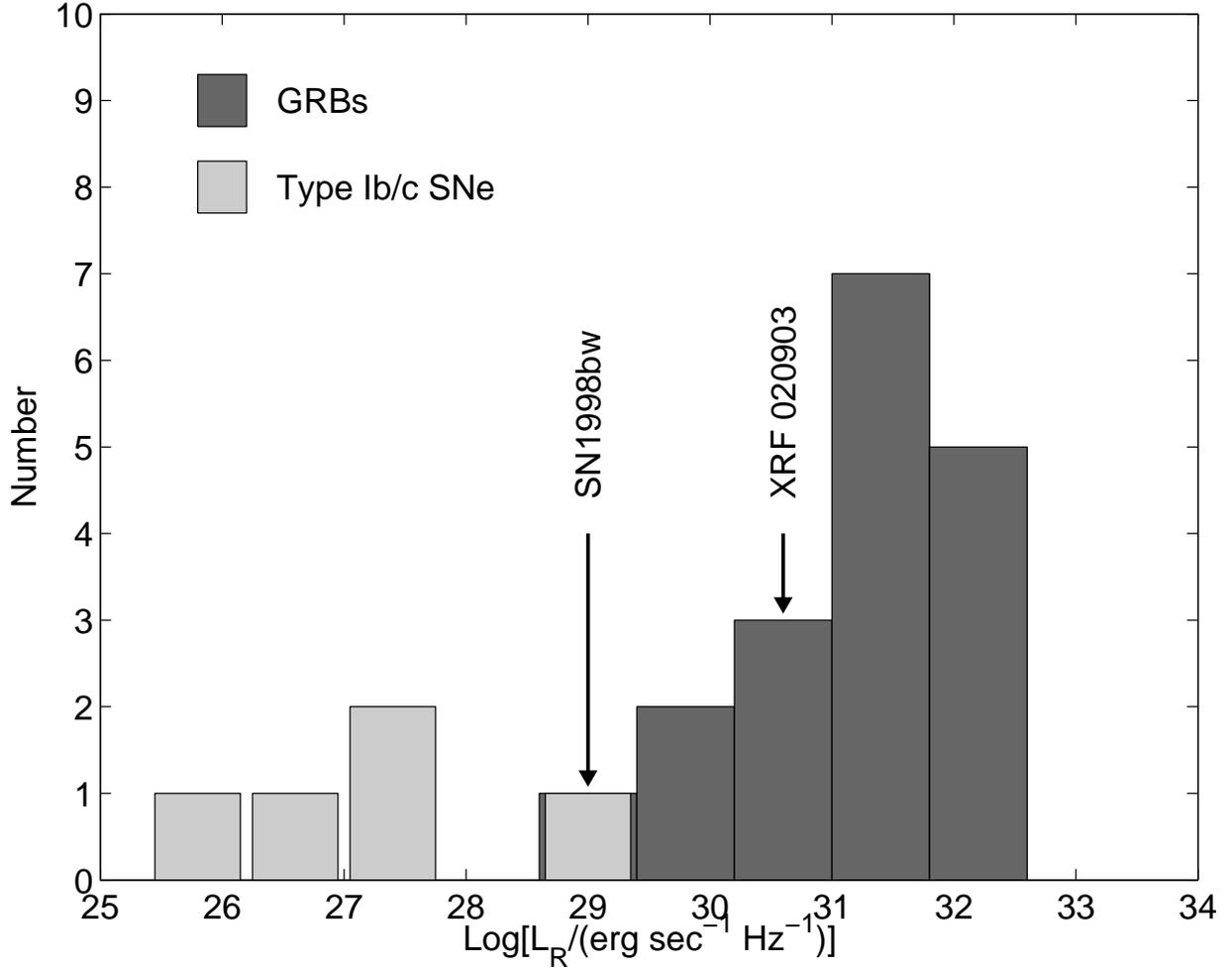}
\caption{The histogram above compares the peak radio luminosity for
various cosmic explosions, including type Ib/c supernovae (spherical
ejecta geometry) as well as typical long-duration GRBs (collimated
ejecta geometry).  Although the energy in the prompt emission of XRF
020903 is $10^2-10^3$ times lower than that observed for typical GRBs,
the total kinetic energy in the afterglow of XRF 020903 is comparable
to that found in typical GRBs afterglows.  In comparison with a larger
sample of cosmic explosions, XRF 020903 is a clear example where there
is less energy coupled to high Lorentz factors.}
\label{fig:XRF020903-GRB-radio-hist}
\end{figure}

\clearpage

\begin{deluxetable}{lrrr}
\tablecaption{Optical Observations of the afterglow of XRF 020903. 
\label{tab:Optical-Photometry}} 
\tablewidth{0pt}
\tablehead{ \colhead{Date} &
\colhead{$\Delta t$} & \colhead{Telescope} & \colhead{R} \\
\colhead{(UT)} & \colhead{(days)} & 
\colhead{} & \colhead{(mag)} \\
} 
\startdata 
2002 Sep 4.32 & 0.9 & Palomar 200'' & 19.23$\pm$0.10 \\
2002 Sep 10.30 & 6.9 & Palomar 200'' & 20.60$\pm$0.10 \\
2002 Sep 28.25 & 24.8 & MDM 1.3 m & 20.80$\pm$0.2 \\
2002 Oct 7.17 & 33.8 & Palomar 200'' & 20.73$\pm$0.17 \\
2003 Jul 2.47 & 302.1 & Palomar 200'' & 21.00$\pm$0.45 \\
\enddata
\end{deluxetable}

\clearpage

\begin{deluxetable}{lrrrrr}
\tablecaption{Radio Observations of the afterglow of XRF 020903. 
\label{tab:Radio-Flux}} 
\tablewidth{0pt}
\tablehead{ \colhead{Date} &
\colhead{$\Delta t$} & \colhead{1.5 GHz} & \colhead{4.9 GHz} &
\colhead{8.5 GHz} & \colhead{22.5 GHz} \\ 
\colhead{(UT)} & \colhead{(days)} & 
\colhead{($\mu$Jy)} & \colhead{($\mu$Jy)} & \colhead{($\mu$Jy)} &
\colhead{($\mu$Jy)} 
} 
\startdata 
2002 Sep 27.22 & 23.8 & ... & ... & 1058$\pm$19 & ... \\
2002 Sep 29.11 & 25.7 & 197$\pm$70 & 552$\pm$42 & 746$\pm$37 & 590$\pm$125 \\
2002 Sep 30.23\tablenotemark{a} & 26.8 & ...  & ... & 892$\pm$166 & ... \\
2002 Oct 2.06 & 28.6 & 210$\pm$91 & 788$\pm$45 & 765$\pm$41 & 0$\pm$270 \\
2002 Oct 10.13 & 36.7 & 294$\pm$91 & 832$\pm$47 & 640$\pm$40 & ...\\
2002 Nov 13.02 & 74.0 & 175$\pm$56 & 380$\pm$38 & 190$\pm$30 & ... \\
2003 Mar 3.71 & 181.3 & ... & 215$\pm$59 & ... & ... \\
2003 Mar 7.72 & 185.3 & ... & 112$\pm$45 & ... & ... \\
2003 Mar 14.74 & 192.3 & ... & 127$\pm$32 & ... & ... \\
2003 Mar 17.69 & 195.3 & ... & 103$\pm$40 & ... & ... \\
2003 Sep 15.22 & 376.8 & 43$\pm$66 & 21$\pm$34 & ... & ... \\
 \enddata

\tablenotetext{a}{VLBA observation}




\end{deluxetable}

\end{document}